\documentclass[a4paper,11pt]{article}
\pdfoutput=1 

\usepackage{jheppub} 

\usepackage[T1]{fontenc} 

\title{\boldmath Ehrenfest's scheme and microstructure for regular-AdS black hole in the extended phase space}

\author[a]{Sen Guo,}
\author[a]{En-Wei Liang}

\affiliation[a]{Guangxi Key Laboratory for Relativistic Astrophysics, School of Physical Science and Technology, Guangxi University, Nanning 530004, People's Republic of China}

\emailAdd{sguophys@st.gxu.edu.cn}
\emailAdd{lew@gxu.edu.cn}

\abstract{The regular (Bardeen)-AdS (BAdS) black hole (BH) in the extended phase space is taken as an example for investigating the BH phase transition grade from both macroscopic and microscopic points of view. The equation of state and thermodynamic quantities of this BH are obtained. It is found that the BAdS BH phase space in the extended phase space should be a second-order phase transition near the critical point by verifying the Ehrenfest's equation, and the possibility of its first-order phase transition can be ruled out by the entropy continuity and the heat capacity mutation. The critical exponents from the microscopic structure are analytically and numerically presented with the Landau continuous phase transition theory by introducing a microscopic order-parameter.
}

\begin{document}
\maketitle
\flushbottom

\section{Introduction}
\label{sec:intro}
Black hole (BH) as a thermodynamic system is an ingenious link between classical thermodynamic and general relativity. Since the Hawking radiation was proposed in 1974, the BH thermodynamic properties have been extensively studied \cite{1,2,3,4}. As a pioneer on the BH phase transition, Davies found that the BH phase transition is similar to the classical thermodynamic system \cite{5}. Hawking and Page found that the BH phase transition in the anti-de Sitter (AdS) space becomes more evident and precise \cite{6}. It is also found that the charged AdS BHs share a similar phase transition as the van der Waals (vdW) liquids \cite{7}. Dolan proposed a scheme by taking the cosmological constant as the thermodynamic pressure in the extended phase space \cite{8}. BH thermodynamic properties, including the holographic phase transition \cite{9,10}, $P-\nu$ critical \cite{11,12,13,14} and Joule-Thomson expansion \cite{15,16,17}, were also investigated by some groups.

The phase transition is divided into the first-order and high-order phase transition using the Gibbs free energy continuity in the classical thermodynamics features. The Clausius-Clapeyron-Ehrenfest's equations present a rational judgment of the phase transition. The first-order phase transition satisfies the Clausius-Clapeyron equation, while the second-order phase transition satisfies the Ehrenfest's equation \cite{18,19}. For a singularity space-time, Banerjee $et.al$ found that the Reissner-Nordstr\"{o}m-AdS/Kerr-AdS BH phase transition is the second-order and the main reason that leads to the second-order phase transition is the BH charge/angular momentum \cite{20,21}. Extensive study in this framework for the singularity space-time in various gravity theoretical models shows that the phase transition grade would be different for various BHs, but the BHs' critical behavior is similar to the vdW system \cite{22,23,24,25,26,27,28,29,30,31}. These results are based on the BHs' specific heat diverges at the critical point.

The property of a thermodynamic system can also be described by macroscopic quantities based on its microscopic structure. The thermodynamic system has phase transition latent heat if the liquid and gas phases coexist, and the different phases correspond to different microstructures. It was proposed that a BH may be full of BH molecules like air molecules for carrying the BH entropy microscopic degree freedoms, and its microstructure is similar to the ordinary thermodynamic system \cite{32}. The molecular density of a BH can be defined from the microstructure point of view as \cite{33}
\begin{equation}\label{1}
m=\frac{1}{\nu}=\frac{1}{2l_{p}^{2}r_{+}},
\end{equation}
where $l_{p}$ is the Planck length ($l_{p}\equiv\sqrt{\hbar G_{N} /c^{3}}$) and $r_{+}$ is the event horizon radius. The phase transition is resulted from the change of the BH molecular density, being similar to the influence of magnetization on ferromagnetic phase transition, the electric potential can lead to the polarization and displacement polarization \cite{33}. By constructing a binary fluid model of BH molecules, Zhao $et.al$ discussed the microstructure phase transition of the Reissner-Nordstr\"{o}m-AdS BH and quintessence charged AdS BH. They found that the BH molecules are similar to the ferromagnetic matter \cite{34}. Basing on the thermodynamic geometry, Zhang $et.al$ found that the Gauss-Bonnet coefficient affects the Gauss-Bonnet BH phase transition microstructure \cite{35}.

Besides of the singularity space-time, the regular space-time is also widely studied. Bardeen obtained the first BH solution without singularity at the origin in 1968 \cite{36}. Ay\'{o}n-Beato and Garc\'{\i}a pointed out that the physical source of the regular BHs might be nonlinear electrodynamics \cite{37}, and the thermodynamic properties of these regular BHs have also been discussed in \cite{38,39,40,41,42}.

The phase transition grade and microstructure of the regular BHs in the extended phase space are still of opening questions. This paper focuses on this issue. We study the BAdS BH phase transition types by assuming a negative cosmological constant of an AdS background as a positive thermodynamic pressure from a macroscopic perspective. Considering the BH molecules and the microstructure, we also analyze the BAdS BH phase transition by using the Landau continuous phase transition theory.

The outlines of this paper are listed as follows. Sec.\ref{sec1} gives a brief review of the BAdS BHs thermodynamic. Sec.\ref{sec2} presents our analysis on the BAdS BH phase transition grade and our calculations for several critical exponents by using the Ehrenfest's equation. The analysis of the BH phase transition from the microstructure perspective with the Landau continuous phase transition theory is presented in Sec.\ref{sec3}. Sec.\ref{sec4} ends up with our conclusions and discussion. For simplicity, we adopt the dimensionlization as $G_{N}=\hbar=\kappa_{B}=c=1$.

\section{Thermodynamic of the BAdS BH}
\label{sec1}

Based on the Einstein-Hilbert action, the line element of the BAdS BH in the extended phase space can be derived as \cite{42,43}
\begin{equation}
\label{2}
ds^{2}=-{f(r)}dt^{2}+{f(r)}^{-1}dr^{2}+r^{2}{d\Omega}^{2},
\end{equation}
where
\begin{equation}
\label{3}
f(r)=1-\frac{2 M r^{2}}{{(q^{2}+r^{2})}^{{3}/{2}}}+\frac{r^{2}}{l^{2}},
\end{equation}
in which $M$ is the mass and $q$ is the magnetic charge of the BH, and $l$ is the positive AdS radius. The event horizon radius $r_h$ of a BH is derived from the largest root of $f(r_{h})=0$. Therefore, we get
\begin{equation}
\label{4}
M=\frac{(l^{2}+{r^2_{h}}){(q^{2}+{r^2_{h}})}^{{3}/{2}}}{2l^{2}{r^2_{h}}}.
\end{equation}

In the extended phase space, the cosmological constant is defined as the system pressure, i.e. $P\equiv {3}/{8\pi l^{2}}$ \cite{44,45}. Thus, the BH mass as a function of the entropy $S$, pressure $P$, and magnetic charge $q$ can be written as
\begin{equation}
\label{5}
M=\frac{(\pi q^2+S)^{3/2}(3+8PS)}{6\sqrt{\pi}S}.
\end{equation}
According to the first law of the BH thermodynamics, the BAdS BH temperature $T$, thermodynamic volume $V$, and electrical potential $\varphi$ is given by
\begin{equation}
\label{6}
T=\frac{\sqrt{q^2+S/\pi}(S+8PS^2-2\pi q^2)}{4S^2},
\end{equation}
\begin{equation}
\label{7}
V=\frac{4\pi}{3}\Big(q^2+S/\pi\Big)^{{3}/{2}},
\end{equation}
\begin{equation}
\label{a}
{\varphi}=\frac{\pi q(3+8PS)\sqrt{q^2+S/\pi}}{2S}.
\end{equation}
From equations ({\ref{6}}) and ({\ref{7}}), we obtain the equation of state of the BAdS BH as
\begin{equation}
\label{8}
P=\frac{T}{2\sqrt{q^2+r^2_{h}}}-\frac{1}{8 \pi r^2_{h}}+\frac{q^{2}}{4\pi {r^4_{h}}}.
\end{equation}
Basing on the critical condition \cite{11}
\begin{equation}
\label{9}
\frac{\partial P}{\partial r_{h}}=0=\frac{{\partial}^{2} P}{{\partial}{r^2_{h}}},
\end{equation}
the critical thermodynamic properties $r_{c}$, $P_{c}$, and $T_c$ of the BAdS BH are
\begin{equation}
\label{10}
r_{c}=\sqrt{2(2+\sqrt{10})}q,~~T_{c}=\frac{25(31+13\sqrt{10})}{432(5+2\sqrt{10})^{3/2}\pi q},~~P_{c}=\frac{5\sqrt{10}-13}{432\pi q^{2}}.
\end{equation}
The constant pressure and the constant volume heat capacities of the BAdS BH are
\begin{equation}
\label{11}
C_{P,q}=T\Big(\frac{dS}{dT}\Big)_{P,q}=\frac{2\pi r_{h}^2 (q^{2}+r_{h}^2)(8\pi P r_{h}^4 -2q^{2}+r_{h}^2)}{8\pi P r_{h}^6+8q^{4}+4q^{2}r_{h}^2-r_{h}^4}
\end{equation}
and
\begin{equation}
\label{12}
C_{V,q}=T\Big(\frac{dS}{dT}\Big)_{V,q}=0.
\end{equation}


\section{Ehrenfest's scheme of the BAdS BH}
\label{sec2}

In this section, we investigate the BAdS BH phase transition grade from a macroscopic point of view. The phase transition grade can be determined with the entropy continuity and heat capacity mutation in the classical thermodynamic. The entropy continuity signifies the first-order phase transition, while the heat capacity mutation and specific heat diverge represent the second-order phase transition. Figure 1 shows the temperature as a function of the entropy and the specific heat as a function of the radius of the BAdS BH. It is found that the entropy is a continuous function of the temperature, but the specific heat shows infinite divergence at the critical point, which is a strong signal for the beginning of the higher-order phase transition.
\begin{figure}[h]
\centering 
\includegraphics[width=.45\textwidth]{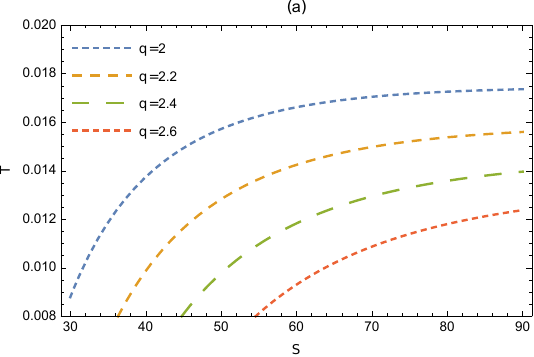}
\includegraphics[width=.45\textwidth]{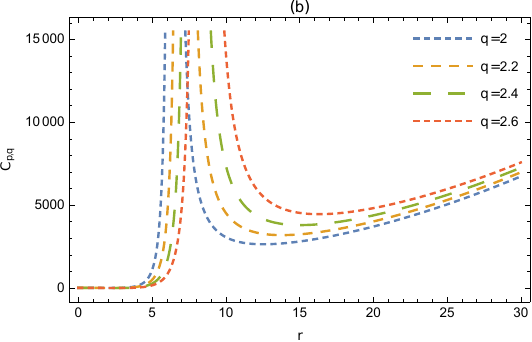}
\caption{\label{fig1} {\em Panel (a)}-- Temperature ($T$) as a function of entropy ($S$) and {\em Panel (b)}-- Specific heat ($C_{P,q}$) as a function of the radius ($r$) of the BAdS BH for $P=P_{c}$ and the magnetic charge taking as $q=2,~2.2,~2.4,~2.6$.}
\end{figure}

Note that more rational judgment for the BAdS BH phase transition classification would base on the Clausius-Clapeyron-Ehrenfest's equations. In the following, we analyze the phase transition phenomena for the BAdS BH in the extended phase space by using the Ehrenfest's scheme. The Ehrenfest's equations are
\begin{equation}
\label{13}
-\Big(\frac{\partial P}{\partial T}\Big)_{S,q}=\frac{C_{P2}-C_{P1}}{ T V (\alpha _2-\alpha _1)}=\frac{\Delta C_{P}}{T V \Delta \alpha},
\end{equation}
and
\begin{equation}
\label{14}
-\Big(\frac{\partial P}{\partial T}\Big)_{V,q}=\frac{\alpha _2-\alpha _2}{\kappa _{T2}-\kappa _{T1}}=\frac{\Delta \alpha}{\Delta \kappa_{T}},
\end{equation}
where $\alpha$ and $\kappa_{T}$ are the expansion coefficient and compressibility coefficient, respectively, i.e.
\begin{equation}
\label{15}
\alpha=-\frac{1}{V}\Big(\frac{\partial V}{\partial T}\Big)_{P,q},~~~\kappa_{T} =-\frac{1}{V}\Big(\frac{\partial V}{\partial P}\Big)_{T,q}.
\end{equation}
According to equations (\ref{6}) and (\ref{7}), the temperature can be re-written as a function of the pressure and volume, i.e.
\begin{eqnarray}
\label{16}
T(P,V)&&=\frac{(V)^{1/3}\big[32 \pi  P q^4-4 q^2 (4\times 6^{2/3} \pi^{1/3} P V^{2/3}+3)\big]}{2^{2/3} \pi ^{4/3} \big[(6/{\pi})^{2/3} V^{2/3}-4 q^2\big]^2}\nonumber\\
&&+\frac{(18/{\pi})^{1/3}\big[12 P V^{2/3}+(6/{\pi})^{1/3}\big]V}{2^{2/3} \pi ^{4/3} \big[(6/{\pi})^{2/3} V^{2/3}-4 q^2\big]^2}.
\end{eqnarray}
Using equations (\ref{15}) and (\ref{16}), the expansion coefficient and compressibility coefficient are re-expressed as
\begin{equation}
\label{17}
\alpha=-\frac{1}{V}\Big(\frac{\partial V}{\partial T}\Big)_{P,q}=-\frac{12 \pi  r_h^6}{\sqrt{q^2+r_h^2} (8 q^4+4 q^2 r_h^2-r_h^4+8 \pi  P r_h^6)},
\end{equation}
and
\begin{equation}
\label{18}
\kappa_{T} =-\frac{1}{V}\Big(\frac{\partial V}{\partial P}\Big)_{T,q}=\frac{24 \pi  r_h^6}{8 \pi  P r_h^6+8 q^4+4 q_h^2 r_h^2-r_h^4}.
\end{equation}
Therefore, the left side of the Ehrenfest's equations is
\begin{equation}
\label{19}
-\Big(\frac{\partial P}{\partial T}\Big)_{S,q}=-\Big(\frac{\partial P}{\partial T}\Big)_{V,q}=-\frac{1}{2 \sqrt{q^2+r_h^2}}.
\end{equation}
Taking $q=2$, we obtain $P=P_{c}=0.0005179$ and $r_h=r_{c}=6.426$ by using equation (\ref{10}). Thus, we have
\begin{equation}
\label{20}
-\Big(\frac{\partial P}{\partial T}\Big)_{S,q}=-\Big(\frac{\partial P}{\partial T}\Big)_{V,q}=-0.074.
\end{equation}
Considering the right side of the Ehrenfest's equations, the specific heat is derived as
\begin{equation}
\label{21}
C_{P1/P2}=\frac{2r_{1/2}^{2}\big[q^2+r_{1/2}^{2}(8P \pi^{2}r_{1/2}^{4}-2\pi q^{2}+\pi r_{1/2}^{2})\big]}{8q^{4}+4q^{2}r_{1/2}^{2}+r_{1/2}^4 (8\pi P r_{1/2}^{2}-1)}.
\end{equation}
In case of $q=2$, we have $T=T_{c}=0.0174276$, $V=V_{c}=1276.86$.  The radius, the specific heats, the expansion coefficients, and the compressibility coefficients on the both sides of the critical point are $r_{1}=6.425$, $r_{2}=6.427$; $C_{P1}=3.43546\times10^{9}$, $C_{P2}=1.62079\times10^{10}$; $\alpha_{1}=2.07862\times10^{9}$, $\alpha_{2}=9.80101\times10^{9}$; $\kappa_{T1}=2.79745\times10^{10}$ and $\kappa_{T2}=1.31941\times10^{11}$. With these results, we have
\begin{equation}
\label{22}
\frac{C_2-C_1}{(\alpha _2-\alpha _1) T V}=\frac{\alpha _2-\alpha _1}{\kappa _{T2}-\kappa _{T1}}=-0.074,
\end{equation}
from the equations (\ref{13}) and (\ref{14}).
\par
This result shows that the BAdS BH phase transition in the extended phase space satisfies the Ehrenfest's equation. The accuracy of our results by using different magnetic charges are reported in table \ref{Tab:Table-1}. To further examine our results, we also plot $\alpha-r$ and $\kappa_{T}-r$ diagrams in figure 2. They are consistent with that reported in table \ref{Tab:Table-1}. These results indicate that the BH phase transition is a second-order phase transition.
\begin{figure}[h]
\centering 
\includegraphics[width=.45\textwidth]{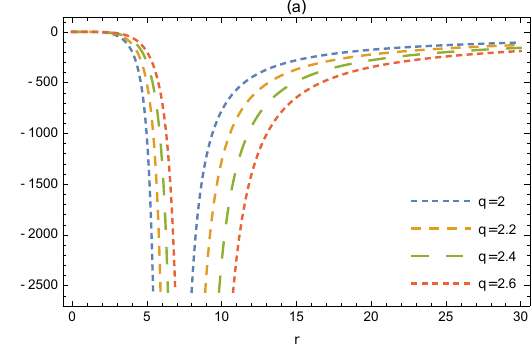}
\includegraphics[width=.45\textwidth]{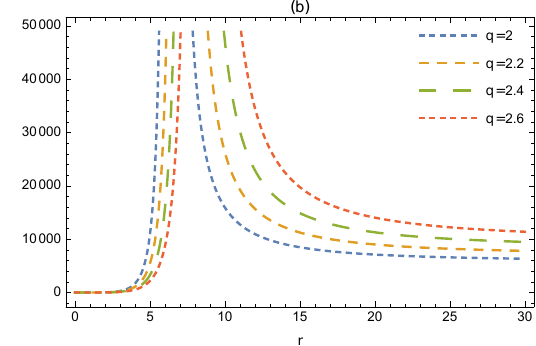}
\caption{\label{fig2} {\em Panel (a)}-- Expansion coefficient ($\alpha$) as a function of the radius ($r$) and {\em Panel (b)}-- Compressibility coefficient ($\kappa_{T}$) as a function of the radius ($r$) of the BAdS BH for $P=P_{c}$ and the magnetic charge taking as $q=2,~2.2,~2.4,~2.6$.}
\end{figure}
\begin{table}
\caption{\bf Different magnetic charges for Ehrenfest's equation}
\label{Tab:Table-1}
\begin{center}
\setlength{\tabcolsep}{1mm}
\linespread{1.5}
\begin{tabular}[t]{|c|c|c|c|c|c|c|c|c|c|}
  \hline
  charge &radius & $\alpha$  &$\kappa_{T}$ &C &$P_{c}$ &$T_{c}$\\
  \hline
  $2.2$  & $r_{c}=7.069$  & & & & $0.000428$ & $0.01584$ \\
         & $r_{1}=7.068$ &$-1.0506\times10^8$ & $1.5555\times10^9$ & $1.910356\times10^8$ & $0.000428$ & $0.01584$\\
         & $r_{2}=7.070$ &$-1.05092\times10^8$ & $1.55628\times10^9$ & $1.9118\times10^8$ & $0.000428$ & $0.01584$\\
   \hline
  $2.4$  & $r_{c}=7.712$  & & & & $0.00036$ & $0.01452$ \\
         & $r_{1}=7.711$ &$-512086$ & $8.27107\times10^6$ & $1.0163\times10^6$ & $0.00036$ & $0.01452$\\
         & $r_{2}=7.713$ &$-511910$ & $8.27118\times10^6$ & $1.0164\times10^6$ & $0.00036$ & $0.01452$\\
   \hline
  $2.6$  & $r_{c}=8.354$  & & & & $0.000306$ & $0.0134$ \\
         & $r_{1}=8.353$ &$392825$ & $-6.8730972\times10^6$ & $-843629$ & $0.000306$ & $0.0134$\\
         & $r_{2}=8.355$ &$392719$ & $-6.8727434\times10^6$ & $-843769$ & $0.000306$ & $0.0134$\\
   \hline
\end{tabular}
\end{center}
\end{table}

The Prigogine-Defay (PD) ratio can also be used to verify the degree of deviation from the standard second-order phase transition. It is unity for the vdW second-order phase transition, while it lies between $2$ to $5$ for the glassy phase transition. Using equations (\ref{17}), (\ref{18}), and (\ref{21}), the PD ratio of the BAdS BH is derived as
\begin{equation}
\label{23}
\Pi=\frac{\Delta C_{p} \Delta \kappa_{T}}{T V (\Delta \alpha)^{2}}=1,
\end{equation}
proving that the BAdS BH phase transition in the extended phase space is a standard second-order phase transition.

The values of the critical exponents are essential characteristics of the second-order phase transition. They are given by
\begin{eqnarray}
& C_{v}=\Big(T\frac{\partial S}{\partial T}\Big)_{V}\propto \mid t \mid^{-\alpha},\label{24}\\
&\eta=V_{l}-V_{s}\propto \mid t \mid^{\beta}, \label{25}\\
&\kappa_{T}=\Big(-\frac{1}{V}\frac{\partial V}{\partial P}\Big)_{T} \propto \mid t \mid^{-\gamma}, \label{26}\\
&(P-P_{c}) \propto (V-V_{c})^{\delta}\label{27}
\end{eqnarray}
where $\alpha$  $\beta$, $\gamma$, and $\delta$ are critical exponents, and $V_{l}$ and $V_{s}$ are the small and large BH volumes.

For the BAdS BH in the extended phase space, the specific heat is zero according to equation (\ref{12}). Thus, the first critical exponent is zero ($\alpha=0$). The other exponents are derived from the reduced thermodynamic variables,
\begin{equation}
\label{28}
p=\frac{P}{P_{c}},~~~v=\frac{\nu}{\nu_{c}},~~~\tau=\frac{T}{T_{c}},
\end{equation}
where $p$, $v$, and $\tau$ are the contrast temperature, the contrast pressure, and the contrast volume, respectively. Expanding them around the critical point by introducing the quantities
\begin{equation}
\label{29}
t=\tau-1, ~~~\omega=v-1,
\end{equation}
we have
\begin{equation}
\label{30}
p\approx0.0987+2.744t+0.936 \omega t-0.052 \omega^{3},
\end{equation}
and
\begin{equation}
\label{31}
dp=-(0.104 \omega^{2}+0.936 t)dw,
\end{equation}
from equations (\ref{8}), (\ref{28}), and (\ref{29}). According to the Maxwell's law of equal area, one can get
\begin{eqnarray}
\label{32}
p&=0.0897+2.744t-0.936 \omega_{l}t-0.052 \omega_{l}^{3}\nonumber\\
&=0.0897+2.744t-0.936 \omega_{s}t-0.052 \omega_{s}^{3},
\end{eqnarray}
and
\begin{equation}
\label{33}
0=\int_{\omega_{s}}^{\omega_{l}}(0.104\omega^{2}+0.936t)\omega d\omega.
\end{equation}
These equations have unique nontrivial solutions, i.e.
\begin{equation}
\label{34}
\omega_{l}=-\omega_{s}\approx4.636\sqrt{-t}.
\end{equation}
Thus, equation (\ref{25}) can be written as
\begin{equation}
\label{35}
\eta=V_{l}-V_{s}=V_{c}(\omega_{l}-\omega_{s})\approx6.5V_{c}\sqrt{-2t}.
\end{equation}
Therefore, the second critical exponent is $\beta=1/2$. Solving equations (\ref{16}), (\ref{26}), and (\ref{30}), the isothermal compressibility coefficient is derived as
\begin{equation}
\label{36}
\kappa_{T}=-\frac{1}{(\omega+1)P_{c}} \frac{d \omega}{d p} \propto \frac{1}{7.02t},
\end{equation}
hence the third critical exponent is $\gamma=1$. For $T=T_{c}$ at $t=0$, the equation (\ref{27}) becomes
\begin{equation}
\label{37}
p-1=0.987-0.052\omega^{3},
\end{equation}
The fourth critical exponent is $\delta=3$. Thus, we obtain the critical exponents for the BAdS BH in the extended phase space are $(\alpha,~\beta,~\gamma,~\delta)$ =(0,~1/2,~1,~3).

\section{Continuous phase transition and microstructure of the BAdS BH}
\label{sec3}
According to Landau's theory, the continuous phase transition is characterized by the order degree and symmetry change of matter. Taking the ferromagnetic matter as an example, the phase transition process is mainly divided into the following stages: \emph{i)} the matter in the lowest energy state at absolute zero with low symmetry, high-order degree, and a non-zero order-parameter; \emph{ii)} the matter in a stage that the molecule thermal movement destroys the ordered orientation of the magnetic molecules with the temperature increases; \emph{iii)} the disordered molecular orientation stage with high symmetry, low-order degree and zero order-parameter as the system temperature are above the critical temperature. In particular, the order-parameter continuously changes from zero to non-zero at the ordered orientation of the magnetic molecules critical point as the temperature decreases.

We propose that the BAdS BH system has similar phase transition behaviors. The thermodynamic characteristics and critical behavior of the charged and non-charged BHs are studied. It is found that the non-charged BH has no phase transition like the vdW system. The charge plays a crucial role in the phase transition, which should be considered in the BH microstructure theory \cite{12,Cai}. The BAdS BH has the mutation potential $\varphi$ at the phase transition point, and it is controlled by its temperature. Hence, we choose the potential $\Psi(T)$ as the order-parameter to analyze the BAdS BH phase transition in the extended phase space from the microscopic view.

Considering the Maxwell's equal area law $P_{0}(V_{2}-V_{1})=\int^{V_{1}}_{V_{2}}PdV$ and equations (\ref{6}), (\ref{7}), and (\ref{8}), one can get
\begin{equation}
\label{a}
T_{0}=\frac{\sqrt{(q^{2}+r_{2}^{2})(q^{2}+r_{2}^{2}x^{2})}(r_{2}^{2}-2q^{2}+2q^{2}x^{4}-r_{2}^{2}x^{4})}{4\pi r_{2}^{4}x^{4}\Big(\sqrt{q^{2}+r_{2}^{2}}-\sqrt{q^{2}+r_{2}^{2}x^{2}}\Big)},
\end{equation}
\begin{eqnarray}
\label{b}
P_{0}&&=\frac{3\Big(q^{2}(\sqrt{q^{2}+r_{2}^{2}}x^{2}-\sqrt{q^{2}+r_{2}^{2}x^{2}})+r_{2}^{2}x^{2}(\sqrt{q^{2}+r_{2}^{2}})\Big)}{A}\nonumber\\
&&+\frac{3\Big(2\pi r_{2}^{2}T(x^{2}-1)-\sqrt{q^{2}r_{2}^{2}q^{2}}\Big)}{A},
\end{eqnarray}
where $A\equiv8\pi r_{2}^{2}x^{2}\Big(q^{2}\sqrt{q^{2}+r_{2}^{2}x^{2}}-\sqrt{q^{2}+r_{2}^{2}}\Big)+r_{2}^{2}\Big(x^{2}\sqrt{q^{2}+r_{2}^{2}x^{2}x^{2}}-\sqrt{q^{2}+r_{2}^{2}x^{2}}\Big)$,
\begin{equation}
\label{c}
r_{2}=q\Big(\frac{x(x^{3}+6x^{2}+6x+1)+\sqrt{B}}{x^{5}}\Big)^{1/3},
\end{equation}
in which $B\equiv x^{2}(x^{6}+12x^{5}+54x^{4}+82x^{2}+54x^{2}+12x+1)$, $x\equiv r_{1}/r_{2}$, and $r_{1}$, $r_{2}$ are the radii of BHs for phase transition points. The case of $x=1$ gives the critical values of $T_{C}$, $P_{C}$ and potential $\varphi_{C}$. If the temperature is fixed and lower than the critical temperature, the electric potentials of any two phases are
\begin{equation}
\label{38}
\varphi_{1}=\frac{q\sqrt{q^2 + r_{1}^2}(3+8P_{1}\pi r_{1}^2)}{2r_{1}^2},~~~\varphi_{2}=\frac{q\sqrt{q^2 + r_{2}^2}(3+8P_{2}\pi r_{2}^2)}{2r_{2}^2}.
\end{equation}
The order-parameter, which is defined as
\begin{equation}
\label{39}
\Psi(T)=\frac{\varphi_{1}-\varphi_{2}}{\varphi_{C}},
\end{equation}
as a function of $T/T_c$ is shown in figure 3. One can observe that the order-parameter is continuous at the critical temperature, which leads to the breaking of time-reversal symmetry spontaneously.

\begin{figure}[tbp]
\centering 
\includegraphics[width=.5\textwidth]{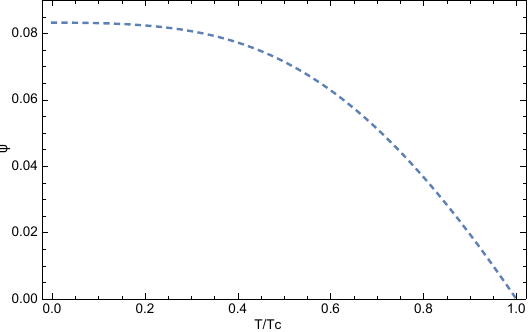}
\caption{\label{fig3} The order-parameter vs reduced temperature plot of the BAdS BH in the extended phase space.}
\end{figure}

It is worth mentioning that the break of the time-reversal symmetry in phase transition in low temperatures for the ferromagnetic matter was investigated, and the spatial rotation symmetry is also broken if the spatial dimension is more than two-dimensional \cite{Yang}. As a high dimensional system, the magnetic potential of the BH equals to the four-vector potential in the dual expression $B_{t}(r_{+})-B_{t}(\infty)$. The gauge transformation of the four-vector potential does not change the difference between the potentials. In the microscopic discussion of BH phase transition, the symmetry breaking caused by the selection of order-parameters is only understood from the perspective of thermodynamic phase space. It is nothing to do with the symmetry of the BH space-time structure.

Similar to the ferromagnetic matter phase transition, we divide the BH phase transition into the following processes. The BH temperature is lower than the critical temperature ($T<T_{C}$) in phase $1$. The BH molecules have a high coefficient phase and low symmetry, high-order degree, and non-zero order-parameter $\Psi_{1}(T)$, which is subjected to strong electric potential $\varphi_{1}$. In phase $2$, the BH molecules have a low coefficient phase. The orientation of the BH molecules resulted from the potential $\varphi_{2}$ is weakened, and the BH molecules have a relatively lower symmetry, higher-order degree, and non-zero order-parameter $\Psi_{2}(T)$ (lower than phase $1$). Then, the thermal movement of molecules destroys the ordered orientation of the BH molecules as the temperature increases. If the temperature is higher than the critical temperature ($T>T_{C}$), the BH molecules orientation appears disorderly, and the system shows a high symmetry, low degree of order. Its order-parameter $\Psi(T)$ is zero.

If the order-parameter is a small quantity around the critical point, the Gibbs function $G(T,\Psi)$ can be expanded by the power of $\Psi(T)$. Since the BH phase transition has symmetry($\varphi \rightleftharpoons -\varphi$), the odd power terms of $\Psi(T)$ are canceled out. Hence, the $G(T,\Psi)$ is expanded as
\begin{equation}
\label{40}
G(T,\Psi)=G_{0}(T)+\frac{1}{2}a(T)\Psi^{2}+\frac{1}{4}b(T)\Psi^{4}+...,
\end{equation}
where the $G_{0}(T)$ is the free energy of $\Psi(T)=0$, and $a,b$ are parameters depending on the system temperature. The phase of $\Psi(T)=0$ can be determined with the minimum condition of stable equilibrium Gibbs function,
\begin{equation}
\label{41}
\frac{\partial G}{\partial \Psi}=\Psi(a+b \Psi^{2})=0,~~~~~\frac{\partial^{2}G}{\partial \Psi^{2}}=a+3b \Psi^{2}>0.
\end{equation}
Solving these equations, one can obtain three solutions, i.e.
\begin{equation}
\label{42}
\Psi=0,~~~~\Psi=\pm\sqrt{-\frac{b}{a}}.
\end{equation}
The solution $\Psi=0$ represents that BH molecules are in a disordered state, which corresponds to the temperature ranges of $T>T_{c}$ and $a>0$. The other two solutions represent the BH molecules are in an ordered state. Substituting the solutions $\Psi=\pm\sqrt{-{b}/{a}}$ into equation (\ref{41}), one can get the system with $T<T_{c}$ and $a<0$. The parameter $a$ is zero at the critical point because the order-parameter $\Psi(T)$ changes from zero to non-zero at this point.

Assuming that $a=a_{0}\Big(\frac{T-T_{c}}{T_{c}}\Big)=a_{0}t (a_{0}>0)$ and $b(T)=b(const)$ around the critical point, it is found that the parameter $a$ is a negative number when $T<T_{c}$, because the order-parameter $\Psi=\pm\sqrt{-b/a}$ is a real number. Hence, we have $b>0$. From the above analyzes, one can obtain
\begin{equation}
\label{43}
\Psi=\left\{
\begin{array}{rcl}
0 ~~~~~~~~~~~~~~~~~~~~~& & {t > 0},\\
\pm(a_0 / b)^{1/2}(-t)^{1/2} & & {t < 0}.
\end{array} \right.
\end{equation}
According to the above equations, the equation (\ref{40}) can be written as
\begin{equation}
\label{44}
G(T,\Psi)=\left\{
\begin{array}{rcl}
G_{0}(T) ~~~~~~~~~~~~~~~~& & {T>T_{C}},\\
G_{0}(T)-\frac{a_{0}^{2}}{4b}\Big(\frac{T-T_{C}}{T_{C}}\Big) & & {T<T_{C}}.
\end{array} \right.
\end{equation}

Since $C=-T(\partial^{2}G /\partial T^{2})$, the difference of the specific heat between two phases at the critical point is
\begin{equation}
\label{45}
C(t\rightarrow -0)-C(t\rightarrow +0)=\frac{a_{0}^{2}}{2b T_{C}}.
\end{equation}
It can be seen from the above formulas, the BH specific heat of the ordered state is larger than that of the disordered state, and the mutation of the specific heat at $t=0$ is limited. Therefore, the first critical exponent $\alpha$ is $0$.

The equation (\ref{43}) shows that the order-parameter jumps at the critical point, indicating that the BAdS BH phase transition is a second-order phase transition. The dependence of the order-parameter $\Psi(T)$ on $t$, which is described by equation (\ref{43}), is similar with equation (\ref{25}). Hence, the second critical exponent $\beta$ is $1/2$.

The BH total differential of the Gibbs free energy at constant pressure is
\begin{equation}
\label{46}
dG=-SdT-\chi qd\Psi,
\end{equation}
where the $\chi$ is a constant, representing the vacuum permeability. One can get
\begin{equation}
\label{47}
-\chi q=\Big(\frac{\partial G}{\partial \Psi}\Big)_{T,P}=a\Psi+b\Psi^{3},
\end{equation}
and
\begin{equation}
\label{48}
-\Big(\frac{\partial \Psi}{\partial q}\Big)_{T,P}=\frac{\chi}{a+3b\Psi^{2}}=\left\{
\begin{array}{rcl}
\frac{\chi}{a_{0}}t^{-1} ~~~~~~~~~~& & {t>0},\\
\frac{\chi}{2a_{0}}(-t)^{-1}~~~~~ & & {t<0}.
\end{array} \right.
\end{equation}
Therefore, by comparing equations (\ref{48}) and (\ref{26}), the third critical exponent $\gamma$ is $1$.

Finally, considering equation (\ref{47}) at the phase transition point $a=0$, one can get
\begin{equation}
\label{49}
q\propto \Psi^{3},
\end{equation}
suggesting that the fourth critical exponent is $\delta=3$. Actually, these exponents are not independent of each other, and they satisfy the following scaling laws
\begin{eqnarray}
\label{50}
\alpha+2\beta+\gamma=2,~~~\alpha+\beta(1+\delta)=2,\nonumber\\
\gamma(1+\delta)=(2-\alpha)(\delta-1),~~~\gamma=\beta(\delta-1).
\end{eqnarray}

\section{Conclusions and discussions}
\label{sec4}

The BAdS BH stands for a space-time without the existence of a singularity. We attempt to investigate the effect of space-time without singularity on the BH phase transition. This paper has investigated the regular BAdS BH phase transition grade in the extended phase space from macroscopic and microscopic perspectives. Firstly, the equation of state and thermodynamic quantities of this BH is derived, and the possibility of the first-order phase transition is removed the entropy continuity and heat capacity mutation. We find that the BAdS BH phase transition is the second-order phase transition by using the Ehrenfest's equation, and the derived the critical exponents are also consistent with the vdW system, which are $(\alpha,~\beta,~\gamma,~\delta)$ =(0,~1/2,~1,~3). It implies that the singularity existence does not affect the occurrence of BH phase transition from the macroscopic view. We get the BAdS BH phase transition as a standard second-order phase transition by calculating the PD ratio.

We compare the BAdS BH phase transition with the ferromagnetic matter phase transition through the Landau continuous phase transition theory. It is found that the regular BAdS BH phase transition is a second-order phase transition from the microscopic view. The BAdS BH temperature determines the BH molecular's order degree, symmetry, and order-parameter, the order-parameter continuously changes from zero to non-zero at the critical point ($T=T_{C}$) with the decrease temperature, and the microstructure of small BH and large BH converge at the critical point. We also found that the critical exponents are consistent with the vdW system and macroscopic perspective, which means that the phase transition of the BAdS BH is a second-order phase transition from various perspectives. The study of the macroscopic thermodynamic properties and microstructure of BHs may provide an important window for exploring the quantum gravity.


\acknowledgments

This work is supported by the National Natural Science Foundation of China (Grant No.11533003, 11851304, and U1731239), by the Guangxi Science Foundation and special funding for Guangxi distinguished professors (2017AD22006).



\end{document}